\documentclass[preprint]{raa}           
\pdfoutput=1
\usepackage{graphicx,times}
\usepackage{natbib}
\usepackage{amssymb,amsmath}
\usepackage{soul}
\bibpunct{(}{)}{;}{a}{}{,}

\graphicspath{{./}}

\usepackage[pagebackref=true]{hyperref}

\begin{document}

   \title{Black hole mass and optical radiation mechanism of the tidal disruption event AT 2023clx}

 \volnopage{ {\bf 20XX} Vol.\ {\bf X} No. {\bf XX}, 000--000}
   \setcounter{page}{1}

   \author{Shiyan Zhong\inst{1}, Xian Xu\inst{1}, Xinlei Chen\inst{1}, Helong Guo\inst{1}, Yuan Fang\inst{1}, Guowang Du\inst{1}, Xiangkun Liu\inst{1}, Xiaowei Liu\inst{1}
   }

   \institute{ South-Western Institute for Astronomy Research, Yunnan University, Kunming, 650500 Yunnan, China; {\it zhongsy@ynu.edu.cn; x.liu@ynu.edu.cn}\\
%        \and
%             Yunnan Astronomical Observatory, National Astronomical Observatories, Chinese Academy of Sciences,
%             Kunmin 650011, China\\
%	\and
%	  Center for Astrophysics, University of Science and Technology of China, Hefei 230026, China\\
%Key Laboratory for Research in Galaxies and Cosmology, The University of Science
%and Technology of China, Chinese Academy of Sciences, Hefei, Anhui, 230026, China\\
\vs \no
   {\small Received 20XX Month Day; accepted 20XX Month Day}
}

\abstract{We present the optical light curves of the tidal disruption event (TDE) AT 2023clx in the declining phase, observed with Mephisto. Combining our light curve with the ASAS-SN and ATLAS data in the rising phase, and fitting the composite multi-band light curves with \texttt{MOSFiT}, we estimate black hole mass of AT 2023clx is between $10^{5.67}$--$10^{5.82}~M_{\odot}$. This event may be caused by either a full disruption of a $0.1~M_{\odot}$ star, or a partial disruption of a $0.99~M_{\odot}$ star, depending on the data adopted for the rising phase. Based on those fit results and the non-detection of soft X-ray photons in the first 90 days, we propose that the observed optical radiation is powered by stream-stream collision. We speculate that the soft X-ray photons may gradually emerge in 100--600 days after the optical peak, when the debris is fully circularized into a compact accretion disk.
}

   \authorrunning{Zhong, Xu, Chen, et al. }            %author_head in even pages
   \titlerunning{BH mass and radiation mechanism of AT 2023clx}  % title_head in odd pages
   \maketitle

%________________________________________________ sections below
% 
\section{Introduction}           %% first-level sections will be auto-capitalized
\label{SECT:Introduction}
A star comes too close to a supermassive black hole (SMBH) shall be disrupted by the overwhelming tidal force from the SMBH, causing a tidal disruption event (TDE). Such an event gives rise to a flare that can last from months to years \cite{Rees1988}. TDEs can occur around black holes of any mass{\footnote{black hole mass must be lower than the Hills mass for a given stellar mass, otherwise the star will be swallowed by the black hole.}}. Therefore, it is a promising probe to detect the SMBHs residing in the center of galaxies and the putative intermediate mass black holes (IMBH) in the dwarf galaxies and globular clusters (See, for example, \citet{Lin+2018NatAs} for a possible TDE occurred around an IMBH).

The critical distance to the SMBH for destroying a star is order of a few to tenths of gravitational radius, $r_{\rm g}=GM_{\rm BH}/c^2$, where $G$, $M_{\rm BH}$ and $c$ are the gravitational constant, black hole mass and speed of light, respectively. Since TDE and the subsequent accretion process happens so close to the SMBH, it is suitable for probing the properties of the SMBH (especially its mass) and the accretion physics. To fulfill this purpose, various tools have been developed and published to the community. Most notable are the Modular Open Source Fitter for Transients (\texttt{MOSFiT})~\citep{MOSFiT,MGR2019}, \texttt{TiDE}~\citep{TiDE} and \texttt{TDEmass}~\citep{TDEmass}. \texttt{MOSFiT} and \texttt{TiDE} work with the multiband light curves (mainly in UV/optical bands), while \texttt{TDEmass} uses the peak bolometric luminosity and the associated effective temperature. These three codes are based on different physical models and assumptions:
\texttt{MOSFiT} adopts a luminosity-dependent photosphere with black body spectrum energy distribution (SED) (detailed in Section~\ref{SUBSEC:LightCurveModel}); 
\texttt{TiDE} adopts an accretion disk and reprocessing layer as the source of the emissions;
\texttt{TDEmass} assumes that the optical/UV photons are emitted from the stream-stream collision region.
Therefore, the three codes may report different values of the SMBH mass and stellar mass. Nevertheless, they are in agreement in terms of order of magnitude (see, e.g., the comparison done by~\cite{KV2023,HvVC2023}). \citet{ZLK2021} has proposed another method to compute the SMBH mass, based on the eccentric accretion disk model developed by \citet{LCA2021}, and take the total emitted energy and the peak bolometric luminosity as the input. \citet{ZLK2021} demonstrated that their results are in well agreement with the SMBH mass estimated from the $M_{\rm BH}$--$\sigma_{*}$ relation. However, the code is currently unavailable to the public.

In the last decade, the number of (candidate and confirmed) TDEs are accumulated at a pace of roughly 2 per year~\citep{Graham2019PASP}, most of them are discovered in the optical band, thanks to the ground based optical surveys. The discovery rate is pushed up to a few dozens per year by the Zwicky Transient Facility (ZTF; \cite{Bellm2019PASP..131a8002B}) and will be further boosted with the upcoming survey facilities, such as the Legacy Survey of Space and Time (LSST) at the Vera Rubin Observatory (VRO) \citep{Ivezic2019ApJ...873..111I}. 

Yunnan University has built a Multi-channel Photometric Survey Telescope (Mephisto)\footnote{http://www.mephisto.ynu.edu.cn/} in September 2022, located at the Lijiang Observatory, Yunnan, China, with longitude $100^{\circ} 01^{\prime} 48^{\prime\prime}$ East, latitude $26^{\circ} 41^{\prime} 42^{\prime\prime}$ North and altitude 3200m. The telescope has a 1.6m primary mirror and covers a field-of-view (FOV) of $2^{\circ}$ in diameter. It is capable of imaging the same FOV in three optical bands simultaneously and delivering real-time, high-quality colors of unprecedented accuracy of surveyed celestial objects. Besides, the $2^{\circ}$ FOV in cooperation with the 3-channel simultaneous exposure, making its survey efficiency competitive to other survey telescopes. Mephisto is currently equipped with two commercial Oxford Instruments/Andor Technology iKon-XXL single-chip CCD cameras for the blue and yellow channels, allowing imaging respectively $uv$ and $gr$ filters. Each camera employs an e2v CCD231-C6 $6144 \times 6160$ sensor with a pixel size of 15 $\mu$m (corresponding to 0.429 arcsec projected on the sky) and covers an area of about a quarter of the full FOV. Mephisto equips 6 optical filters, implemented in blue (\textit{uv}), yellow (\textit{gr}), and red (\textit{iz}) channels.

In this paper, we present the optical light curve of AT 2023clx, a TDE that took place in the nucleus of the galaxy NGC 3799, obtained with Mephisto. There is no previous information about the central SMBH mass of NGC 3799, except for the value recently reported by \cite{ZJW2023} via the empirical relationship between BH mass and galaxy mass. Hence, AT 2023clx provided a unique chance to measure the central SMBH mass in NGC 3799. The observations and data reduction process are described in Section~\ref{SECT:Data}. Then we briefly introduce the light curve modeling of \texttt{MOSFiT}, which is used in this work to extract the physical parameters of AT 2023clx (Section~\ref{SECT:Fitting_model}). In Section~\ref{SECT:Result_Discussion}, we present the fitting results (Section~\ref{SUBSEC:results}). Based on the fitted black hole mass, stellar mass and penetration factor, we explain that the observed optical emissions are powered by stream-stream collision (Section~\ref{SUBSEC:energy_source}), and accordingly the reason for the non-detection of soft X-ray photon during the observed optical flare stage (Section~\ref{SUBSEC:X_ray}). Finally, we summarize our work in Section~\ref{SECT:Summary}.

%%%%%%%%%%%%%%%%%%%%%%%%%%%%%%%%%%%%%%%%%%%%%%%%%%%%%%%%%%%%%%
%%%%%%%%%%%%%%%%%%%%%%%%%%%%%%%%%%%%%%%%%%%%%%%%%%%%%%%%%%%%%%
\section{Observation Data}
\label{SECT:Data}

\subsection{Mephisto Data}
\label{SUBSECT:MEPHSITO_LC}
AT 2023clx was first detected by the All Sky Automated Survey for SuperNovae (ASAS-SN; \cite{SPG2014}) on February 22, 2023. We did the first follow-up observation in the $u$, $v$ band on February 24. Then the observation was interrupted for 2 days due to the installation of yellow channel camera. \cite{Taguchi2023} took an optical spectrum on February 26, the bluish continuum and broad Balmer emission lines led the authors to suspect that this event might be a TDE. After the completion of the yellow channel camera installation, we immediately resumed the follow-up observations. However, the yellow channel camera suffered from some hardware problems and was replaced with the red channel camera on March 10. Since then, Mephisto worked in two-channel mode, i.e., only the blue and yellow channel cameras were operating. Whenever the weather and observation conditions permitted, we did exposures in the \textit{uvgr} bands on the same night. The observation campaign ended in June, when the rainy season began.

The optical light curve is obtained by doing point spread function (PSF) photometry on the difference image. As a first step, one should select reference images from the observational database. Then the image subtraction between the PSF-convolved science and reference images and the PSF photometry on the difference images are performed with Python package \texttt{Photutils}. The corresponding PSF models are constructed with \texttt{PSFEx}, who uses the star profiles selected from the science and reference images, respectively.

Mephisto is a newly built telescope, so this is the first time we take images of NGC 3799, which causes some problem in selecting the reference images. Since the TDE fades with time, we take the latest images taken in the observation campaign as the reference images ($ug$: June 4; $vr$: June 1). However, the TDE is not completely faded away in these reference images, and the difference flux $f_{\rm diff}$ obtained from PSF photometry is lower than the true flux. The intrinsic flux of the TDE in the reference image, $f_{\rm ref}$, should be added up to $f_{\rm diff}$ to get the real flux. To estimate $f_{\rm ref}$ in the reference images in the $uvgr$ bands, we use the bolometric luminosity ($L_{\rm bol} \simeq 10^{41.5}~\rm{erg/s}$) and effective temperature ($T_{\rm eff}\simeq 12500~\rm{K}$) near MJD 60100 reported by \cite{ZJW2023}.

Finally the multi-band light curves are corrected for galactic extinction with $E(B-V)=0.027$ mag \citep{SFD1998ApJ,SF2011ApJ}. Hence the galactic extinction in each bands are $0.133$ mag (\textit{u} band), $0.124$ mag (\textit{v} band), $0.087$ mag (\textit{g} band), and $0.072$ mag (\textit{r} band).

\subsection{Pre-peak Data}
\label{SUBSECT:LC_OTHERS}

The rising part of the light curve is critical in constraining the SMBH mass \citep{MGR2019}. Although we started observing AT 2023clx immediately after receiving the notification from TNS, the rising part of the light curve is still missed. Fortunately, there were pre-peak observations from the ASAS-SN and the Asteroid Terrestrial Impact Last Alert System (ATLAS; \cite{TDH2018PASP,SSY2020PASP}). We retrieved the host-subtracted $g$-band light curve from the ASAS-SN Sky Patrol photometry pipeline \citep{SPG2014,KSS2017}, and the $c$, $o$-band light curve from the ATLAS forced photometry service. For the purpose of constraining the starting time and rising part of the light curve, we only use the ASAS-SN and ATLAS data before MJD 60000.

The full multiband light curves of AT 2023clx are presented in Figure~\ref{fig:lightcurve}, together with the mock light curves generated from the fitting results (detailed in Section~\ref{SUBSEC:results}). At first, we use both the available ASAS-SN and ATLAS data, but find some inconsistency in the apparent peak time between them. Then we made four groups of composite light curves, each has different data sources of the rising part and these groups were named as: AS (ASAS-SN $g$ band only), ATc (ATLAS $c$ band only), AS+ATc (ASAS-SN $g$ band $+$ ATLAS $c$ band), and AS+ATco (ASAS-SN $g$ band $+$ ATLAS $c$, $o$ bands).

\subsection{Historical Variability in the Nuclear Region of NGC 3799}
\label{SUBSECT:ZTF_LC}

NGC 3799, the host galaxy of AT 2023clx, is classified as a LINER galaxy\footnote{http://simbad.harvard.edu/simbad/sim-id?Ident=NGC+3799\&submit=submit+id}, thus the nuclei may exhibit some low-level activities. The $g$, $r$-band light curves presented by the ALeRCE ZTF Explorer have many spikes in the past. By eye inspection on the scientific images available on the NASA/IPAC Infrared Science Archive\footnote{https://irsa.ipac.caltech.edu/cgi-bin/Gator/nph-dd} (IRSA), we found that the quality of the images correspond to these spikes are usually poor. To exclude the possibility that AT 2023clx is an AGN flare in nature, we retrieve the pre-flare ZTF $g$ and $r$ band images from IRSA. These ZTF images cover roughly 5 years before AT 2023clx and are selected base on the following criteria: airmass less than $1.5$ and limiting magnitude greater than $20~\rm{mag}$.

\begin{figure}
    \centering
    \includegraphics[width=1\linewidth]{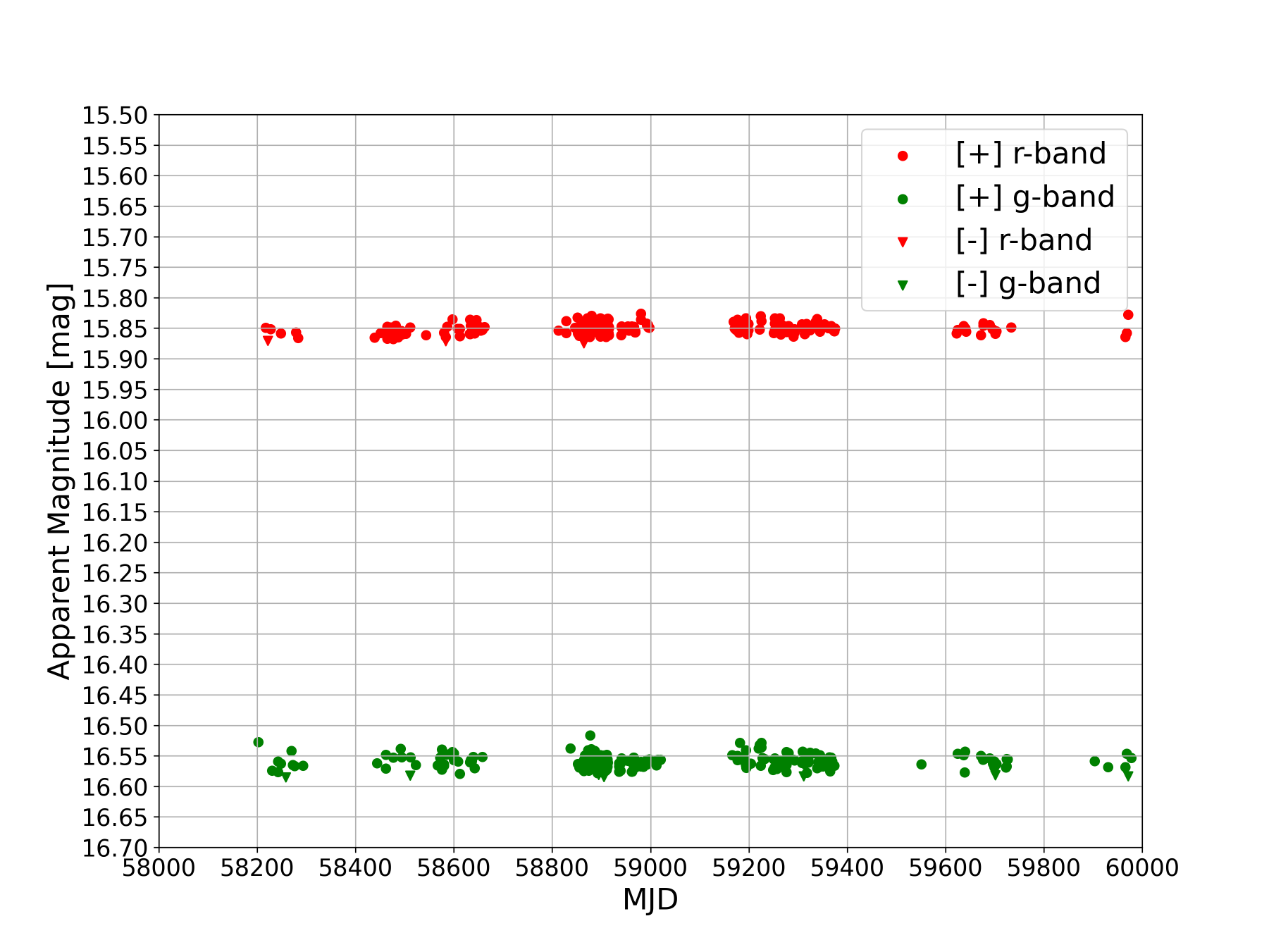}
    \caption{History brightness variation of the nucleus of NGC 3799 in the ZTF $g$ and $r$ bands. The apparent magnitude is computed by $m_{\rm band} = -2.5\log(f_{\rm diff}+f_{\rm base})+ZP$, where $ZP$ is the photometric zero-point in the reference image, $f_{\rm base}$ is the flux of the nuclei obtained from aperture photometry on the reference image, $f_{\rm diff}$ is the flux obtained from PSF photometry on the difference image, respectively. The plus (minus) sign indicates that $f_{\rm diff}$ is positive (negative).
    }
    \label{fig:ZTF_LC}
\end{figure}

The earliest images available in the database are used as the reference images in each ZTF band. Then the difference fluxes $f_{\rm diff}$ in the later exposures relative to the reference image are measured by performing PSF photometry on the difference images. The flux of the nuclei in the reference images $f_{\rm base}$ are obtained via aperture photometry with a diameter of $3$ arcsec. Figure~\ref{fig:ZTF_LC} shows the historical light curve of the nuclei of NGC 3799, with the apparent magnitude computed by $m_{\rm band} = -2.5\log(f_{\rm diff}+f_{\rm base})+ZP$, where $ZP$ is the photometric zero-point recorded in the header of the reference FITS file. The temporal variations of $g$ and $r$ magnitude are smaller than $0.05~\rm{mag}$ in the last five years. For comparison, the brightness of the nuclear region increased maximally $0.3~\rm{mag}$ above the quiescent level during the flare. Thus AT 2023clx is unlikely caused and affected by AGN activity.

% Figure 1

%%%%%%%%%%%%%%%%%%%%%%%%%%%%%%%%%%%%%%%%%%%%%%%%%%%%%%%%%%%%%%%%
%%%%%%%%%%%%%%%%%%%%%%%%%%%%%%%%%%%%%%%%%%%%%%%%%%%%%%%%%%%%%%%%
\section{Light Curve Fitting}
\label{SECT:Fitting_model}

Many valuable information are encoded in the light curve of TDE. In particular, we are interested in the mass of the disrupting SMBH and the mass of the disrupted star.
As mentioned in Section~\ref{SECT:Introduction}, there are three open source software that can tell us the black hole mass from the light curve. \texttt{TDEmass} requires the peak luminosity and the associated effective temperature, unfortunately we did not catch the peak of the light curve. We turn to the \texttt{MOSFiT}, which generates Monte Carlo ensembles of semi-analytical light curve, fit them to the observed data and obtain the associated Bayesian parameter posteriors. It has been used in many studies \citep{MGR2019,Gomez+2020,Nicholl+2020MNRAS,MR2021ApJ,HvVC2023,KV2023}.

\subsection{Light Curve Modeling in \texttt{MOSFiT}}
\label{SUBSEC:LightCurveModel}
In this subsection, we briefly describe the key parameters involved in the light curve model, more details can be found in \cite{MGR2019}.

The modeling starts with mass fallback rate $\dot{M}_{\rm fb}(t)$ measured from hydrodynamic simulation \citep{GRR2013}. In this model, the mass fallback rate are determined by three key parameters: the SMBH mass $M_{\rm BH}$, the stellar mass $m_{*}$ and the penetration factor $\beta$. \cite{GRR2013} adopted two polytropic stellar models: $\gamma=4/3$ (suitable for $1~M_{\odot}\leq m_{*} \leq 15~M_{\odot}$) and $\gamma=5/3$ (suitable for $m_{*} \leq 0.3~M_{\odot}$ and $m_{*} \geq 22~M_{\odot}$). In the rest parts of the mass ranges, hybrid fallback function is adopted and it is constructed as a linear combination of the $\gamma=4/3$ and $\gamma=5/3$ fallback functions, the fractional contributions from the two $\gamma$ models are determined by the stellar mass.

The penetration factor $\beta$ is defined as the ratio between tidal radius $r_{\rm t}$ and the pericenter distance $r_{\rm p}$. For simplicity, the tidal radius is computed by $r_{\rm t}=r_{*}(M_{\rm BH}/m_{*})^{1/3}$, where $r_{*}$ is the stellar radius. One can regard $\beta$ as a proxy of the strength of the tidal field, i.e., higher $\beta$ means the orbital pericenter is closer to the event horizon and more stellar mass shall be stripped from the star. Hence the amount of stripped mass depends on $\beta$. \cite{GRR2013} found that the mass stripping starts at $\beta_{\rm p}$ and full disruption happens at $\beta_{\rm d}$. Accordingly, TDE can be divided into two catagories: parital ($\beta_{\rm p}\leq\beta<\beta_{\rm d}$) and full ($\beta \geq \beta_{\rm d}$) TDE. Note, $\beta_{\rm p}$ and $\beta_{\rm d}$ take different values in different stellar models. To avoid confusion in interpreting the results, \texttt{MOSFiT} maps the conventional $\beta$ to the scaled penetration factor $b$: $b=0$ means no disruption, $0<b<1$ means the star is partially disrupted, $b\geq 1$ means the star is completely disrupted, and $b=2$ corresponds to the maximum $\beta$ that is simulated by \cite{GRR2013}.

The accretion rate $\dot{M}_{\rm acc}(t)$ that actually powers the luminosity should defer from the fallback rate, due to the viscous delay in the accretion disk. In \texttt{MOSFiT}, this process is controlled by the viscous timescale $t_{\nu}$. Increasing $t_{\nu}$ could prolong the rising and declining timescales of the light curve, and suppress the peak luminosity. Note, only in the limit of $t_{\nu} \rightarrow 0$ shall the accretion rate equal the mass fallback rate. The exact formula for transforming $\dot{M}_{\rm fb}(t)$ to $\dot{M}_{\rm acc}(t)$ can be found in \cite{MGR2019}.

The bolometric luminosity is computed by $L_{\rm bol}(t) = \eta \dot{M}_{\rm acc}(t) c^2$, where $\eta$ is the radiation efficiency. The range of $\eta$ is set between $10^{-4}$ and $0.4$, the upper limit is the radiation efficiency for the maximally spinning black hole, while the lower limit comes from minimum value achievable for the eccentric accretion disk model \citep{ZLK2021}.

The optical spectrum energy distribution (SED) of TDEs is generally described by black body radiation with effective temperature $T_{\rm eff}$. The total bolometric luminosity is the product of the area of the emitting surface, $\alpha R_{\rm ph}^2$, and the energy flux per unit area $\sigma_{\rm SB} T_{\rm eff}^4$, where $R_{\rm ph}$ is the so called photosphere radius, $\sigma_{\rm SB}$ is the Stefan-Boltzmann constant and $\alpha$ is a dimensionless geometric factor (in the case of spherical photosphere, $\alpha = 4\pi$; there is also other choice, e.g. in the model of \cite{TDEmass}, $\alpha = 2\pi$). After some simple algebra, the expression for the effective temperature reads

\begin{equation}
T_{\rm eff} = \left ( \frac{L_{\rm bol}}
   {4\pi \sigma_{\rm SB} R_{\rm ph}^2} \right )^\frac{1}{4}.
\label{Eq:T_eff}
\end{equation}
\noindent
Then the observed SED is $F(\lambda)=B_{T_{\rm eff}}(\lambda) R_{\rm ph}^2 / D_{\rm L}^2$, where $B_{T_{\rm eff}}(\lambda)$ is the Planck function and $D_{\rm L}$ is the luminosity distance.

Many observations of TDEs have found that the effective temperature varies little near the peak and tend to increase at late times. In order to model this SED behavior in \texttt{MOSFiT}, a power-law scaling relation between the $R_{\rm ph}$ and $L_{\rm bol}$ is adopted, i.e., $R_{\rm ph} = R_{\rm ph0} a_{\rm p}(L_{\rm bol}/L_{\rm Edd})^l$, where $R_{\rm ph0}$ is a normalization of the photosphere radius, $a_{\rm p}$ is the semimajor axis of the material corresponding to the maximum fallback rate, and $L_{\rm Edd}$ is the Eddington luminosity. Substitute this relation into equation~\ref{Eq:T_eff} results in $T_{\rm eff}\propto L_{\rm bol}^{(1-2l)/4}$, hence in the special case of $l=1/2$, the effective temperature will not vary with time. \cite{Jiang_2016ApJ} studied the stream-stream collision process, they also found power-law relation between $R_{\rm ph}$ and the rate of mass injection into the collision region, with power-law index close to 1.

In order to compare the mock light curve generated by \texttt{MOSFiT} TDE module to the observations. A conversion from bolometric luminosity to the AB magnitude in different bands are performed. This is done with the $F(\lambda)$ and the filter transmission function $T(\lambda)$, using the following equation

\begin{equation}
m_{\rm AB} = -2.5\log \int \lambda F(\lambda) T(\lambda) d\lambda
+2.5\log \int \frac{T(\lambda)}{\lambda} d\lambda -2.408.
\label{Eq:AB_mag}
\end{equation}
\noindent
Finally, a logarithmic likelihood score $\log p$ is computed for the mock light curve (see equation (3) of~\cite{MOSFiT}). This score is used to assess the goodness of fitting, in the sense that the higher the score, the better the mock light curve matches the observation.

%%%%%%%%%%%%%%%%%%%%%%%%%%%%%%%%%%%%%%%%%%%%%%%%%%%%%%
\subsection{Parameter Settings}
\label{SUBSEC:parameter}

The ranges of the aforementioned fitting parameters and the types of their prior distributions are listed in Table~\ref{table-paras}. The rest of fitting parameters use their default settings. To initialize the Markov Chain Monte Carlo (MCMC) fitting procedure, the model parameters are randomly drawn from the prior distributions. In order to accelerate the convergence of the fitting procedure, we narrowed the ranges of the priors for the SMBH mass and stellar mass. \cite{ZJW2023} reported the SMBH in the nuclei of NGC 3799 has a mass of $10^{6.26 \pm 0.28}~M_{\odot}$, derived from the empirical relationship between the central BH mass and total galaxy mass \citep{Reines_Volonteri_2015ApJ}. We thus made the searching area roughly $1$ dex above and below this value, i.e., $5\leq\log(M_{\rm BH}/M_{\odot})\leq 7$. The spectrum of the nuclei taken before the TDE indicate that the stellar population is relatively old, hence we restrict the searching range of stellar mass to low mass range, i.e., $0.08 \leq m_*/M_{\odot} \leq 3$ and the priors are drawn from the Kroupa initial mass function \citep{Kroupa+1993MNRAS}. We also constrained the start time of TDE to be within 30 days before the peak, which is determined from the last non-detection in the ZTF survey. The other fitting parameters take the default ranges and distribution types shipped with \texttt{MOSFiT}. The luminosity distance is set to $47.8$ Mpc, derived from the redshift of NGC 3799, $z=0.011$, in a flat universe ($H_0=69.6~\rm{km/Mpc}$, $\Omega_{\rm M}=0.286$). The fitting procedure stops when the potential scale reduction factor (PSRF) value drops below 1.1, which is the default value in \texttt{MOSFiT} and indicates that the Monte Carlo chain has converged to the target distribution.

\texttt{MOSFiT} uses ``Watanabe-Akaike information criteria (WAIC)" to assess the goodness of fitting, which is defined as: WAIC $= \overline{\log(p)} - var(\log(p))$, where $\overline{\log{p}}$ is the mean of the logarithmic likelihood score and $var(\log{p})$ is the variance, using the samples from the ensemble. According to the definition, the models with higher WAIC values are more appropriate for the observed data. We also present the WAIC values for the four models in Table~\ref{table-results}.

% Figure 2
\begin{figure}
    %\centering
    \includegraphics[width=0.49\linewidth]{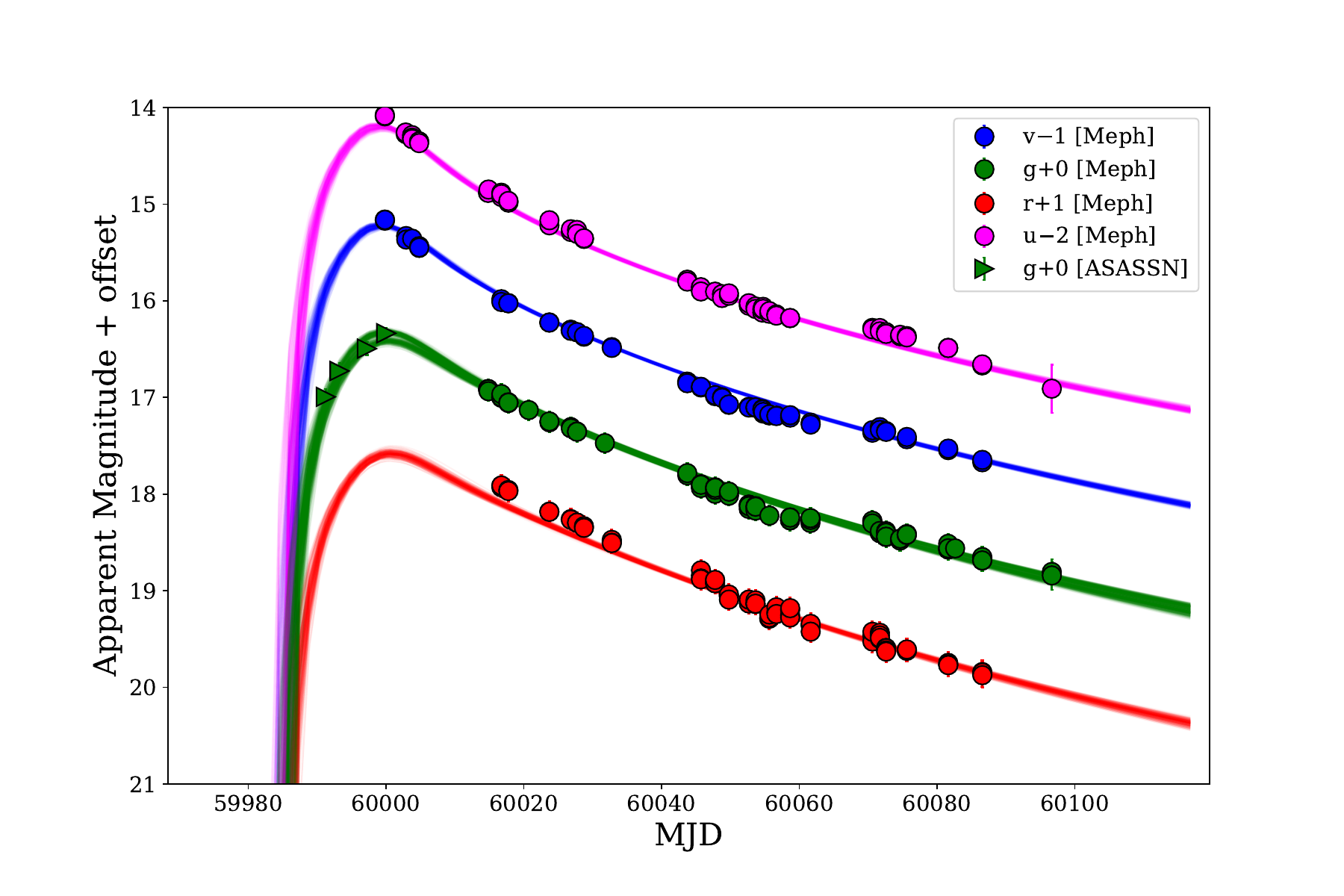}
    \includegraphics[width=0.49\linewidth]{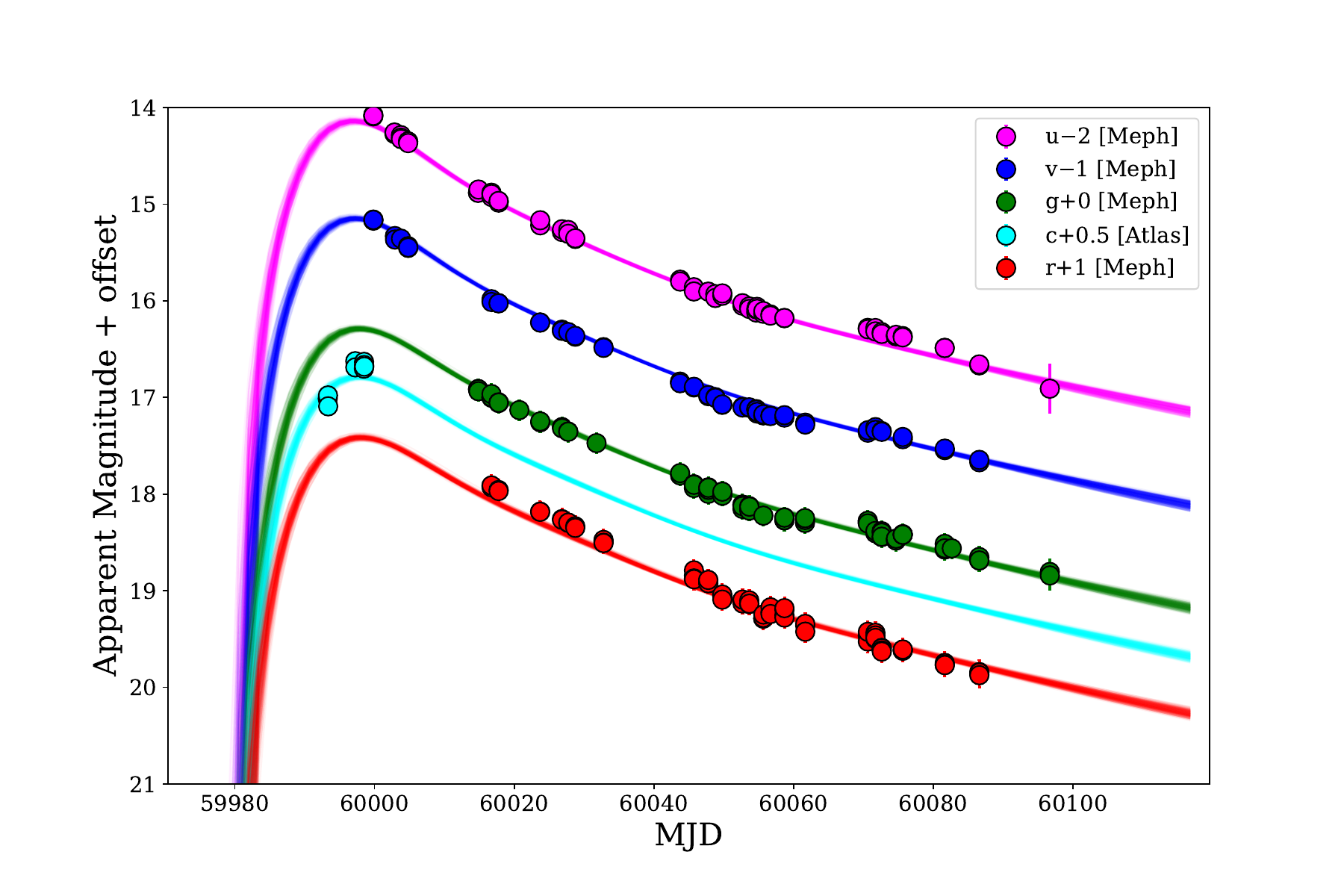}
    \includegraphics[width=0.49\linewidth]{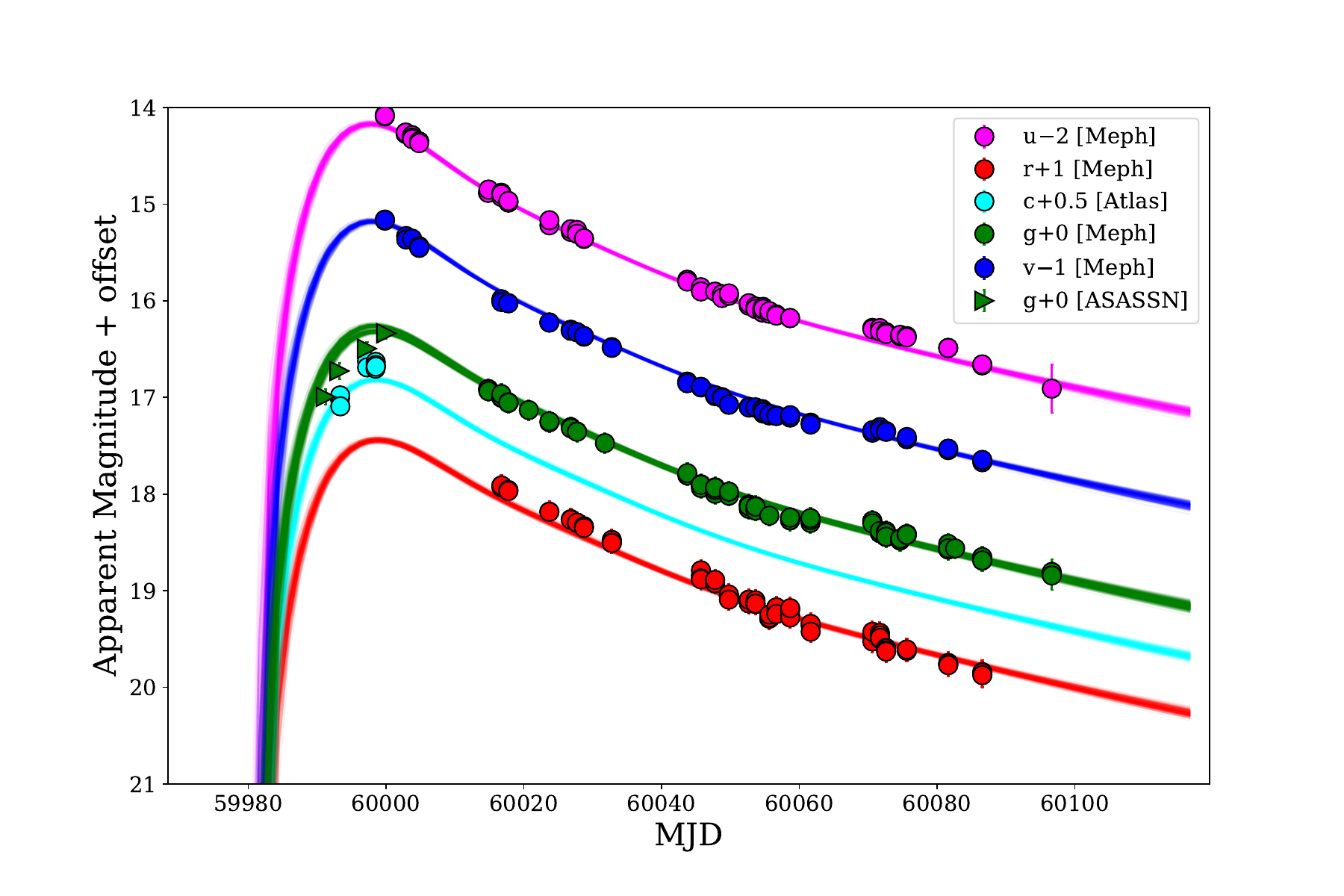}
    \includegraphics[width=0.49\linewidth]{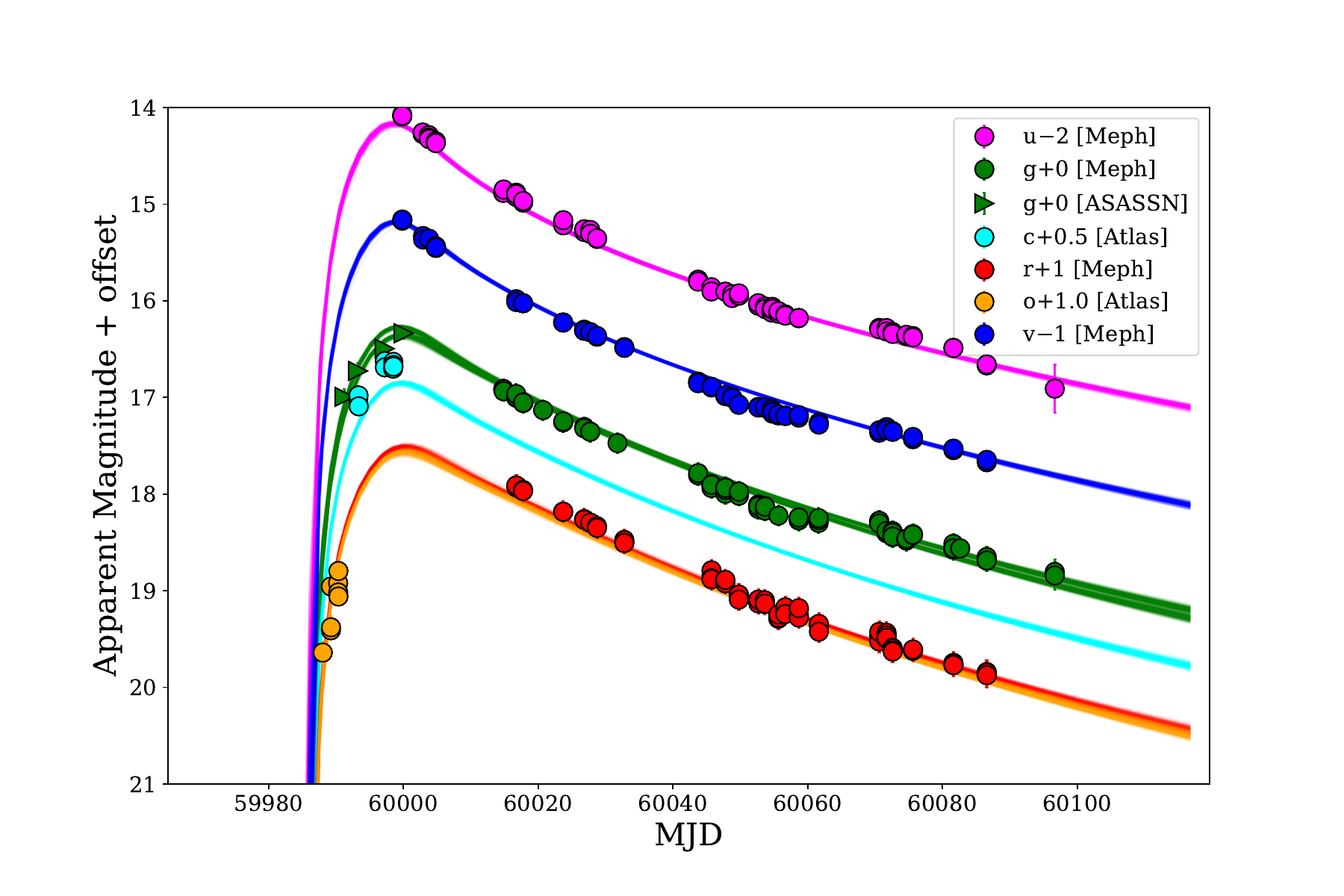}
    \caption{Multiband light curve of AT 2023clx, taken from Mephisto \textit{uvgr} bands, ASAS-SN $g$ band and ATLAS $c$ bands. The four panels use different pre-peak data points: ASASS-SN $g$ band only (AS, top left), ATLAS $c$ band only (ATc, top right), ASAS-SN $g$ band $+$ ATLAS $c$ band (AS+ATc, bottom left), ASAS-SN $g$ band $+$ ATLAS $c$, $o$ band (AS+ATco, bottom right). Superpositioned are the mock light curves generated by \texttt{MOSFiT}.
    }
    \label{fig:lightcurve}
\end{figure}

% Figure 3
\begin{figure}
    \centering
    \includegraphics[width=0.49\linewidth]{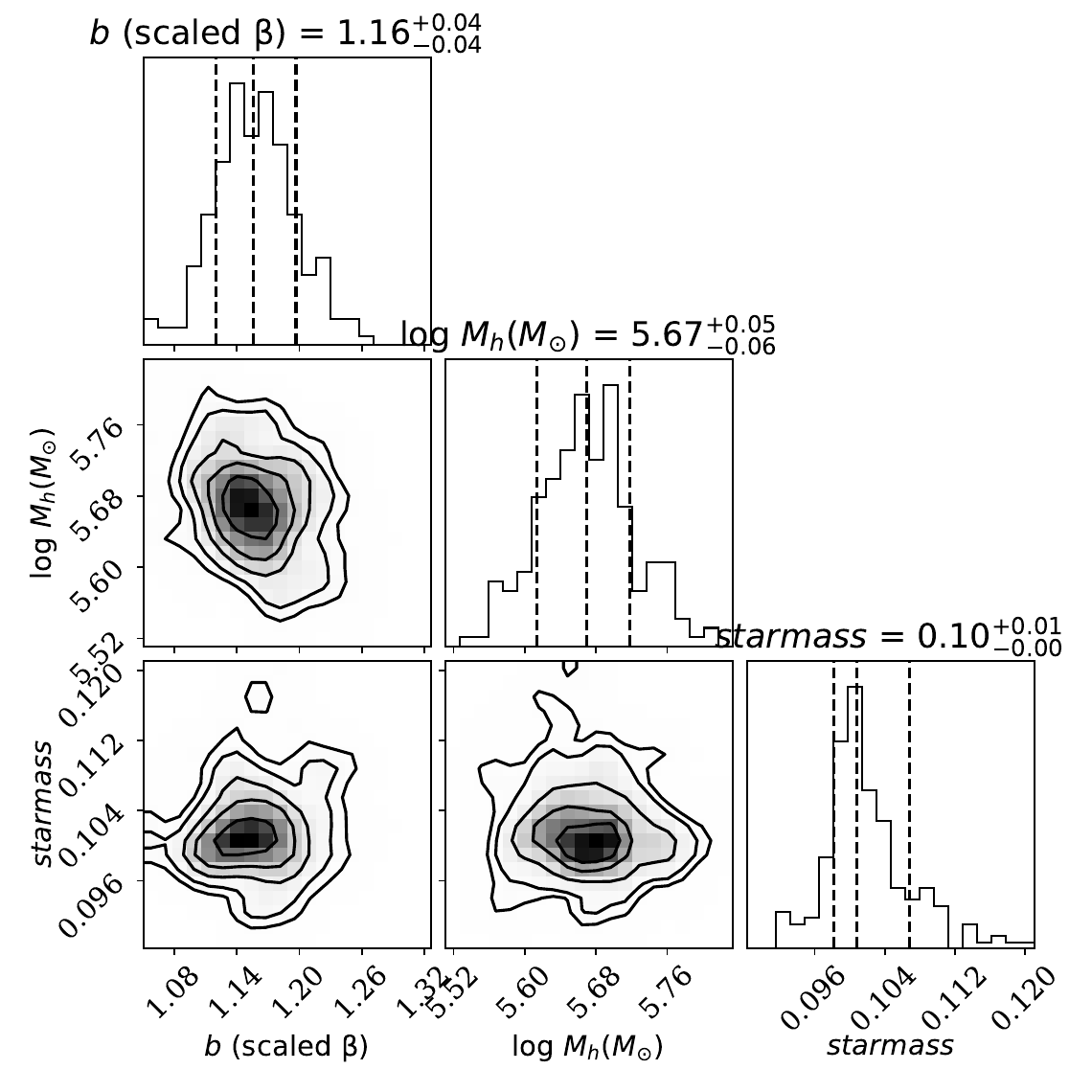}
    \includegraphics[width=0.49\linewidth]{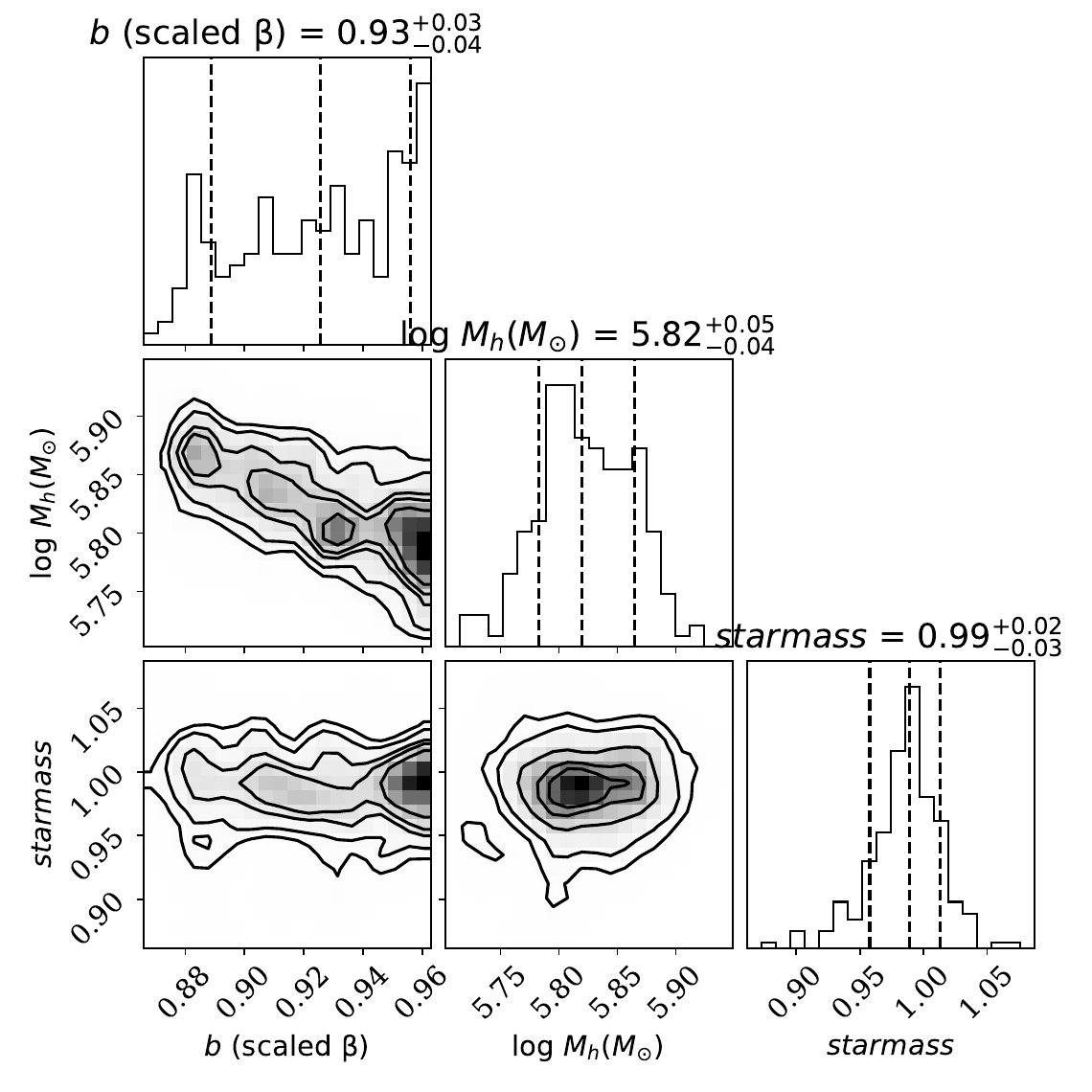}
    \includegraphics[width=0.49\linewidth]{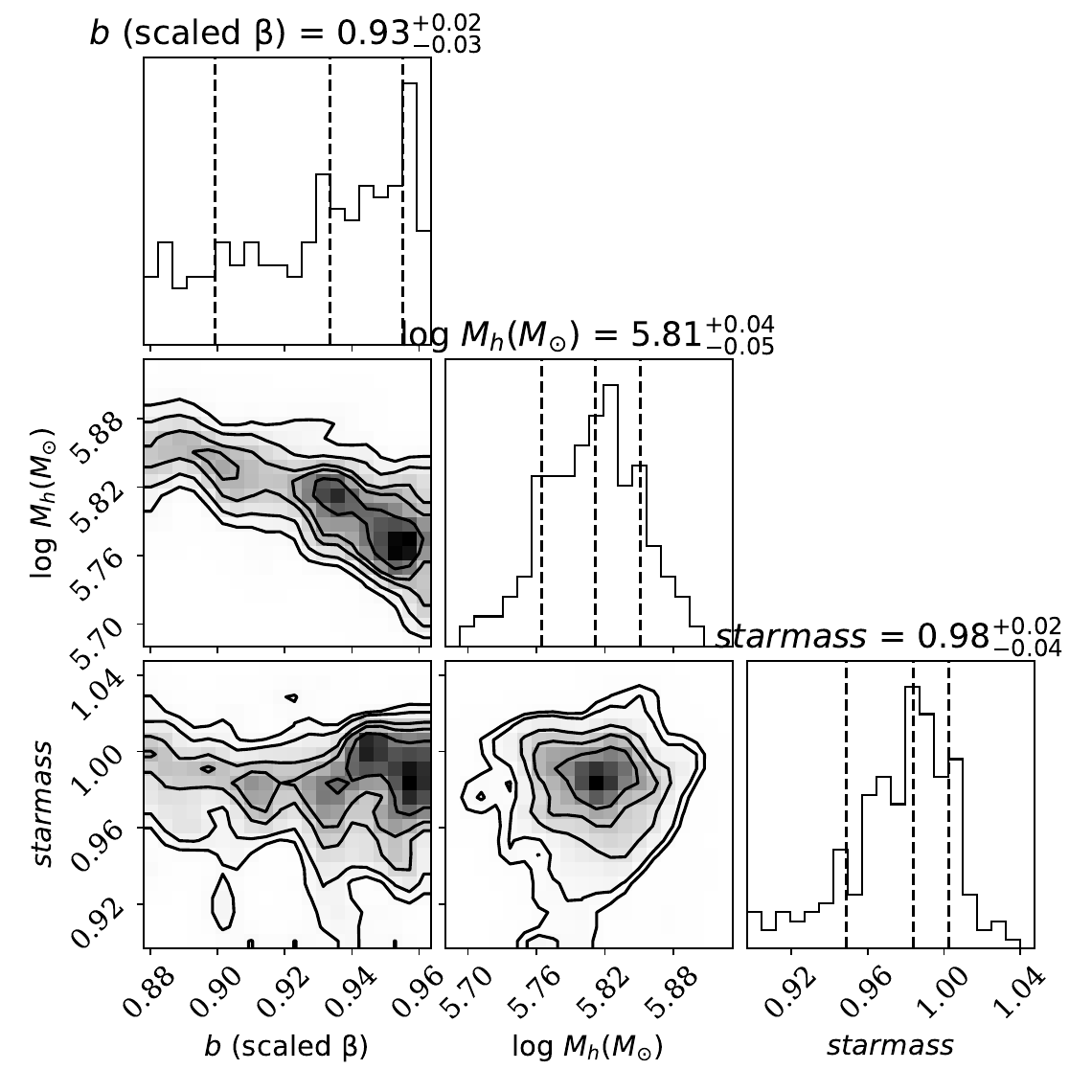}
    \includegraphics[width=0.49\linewidth]{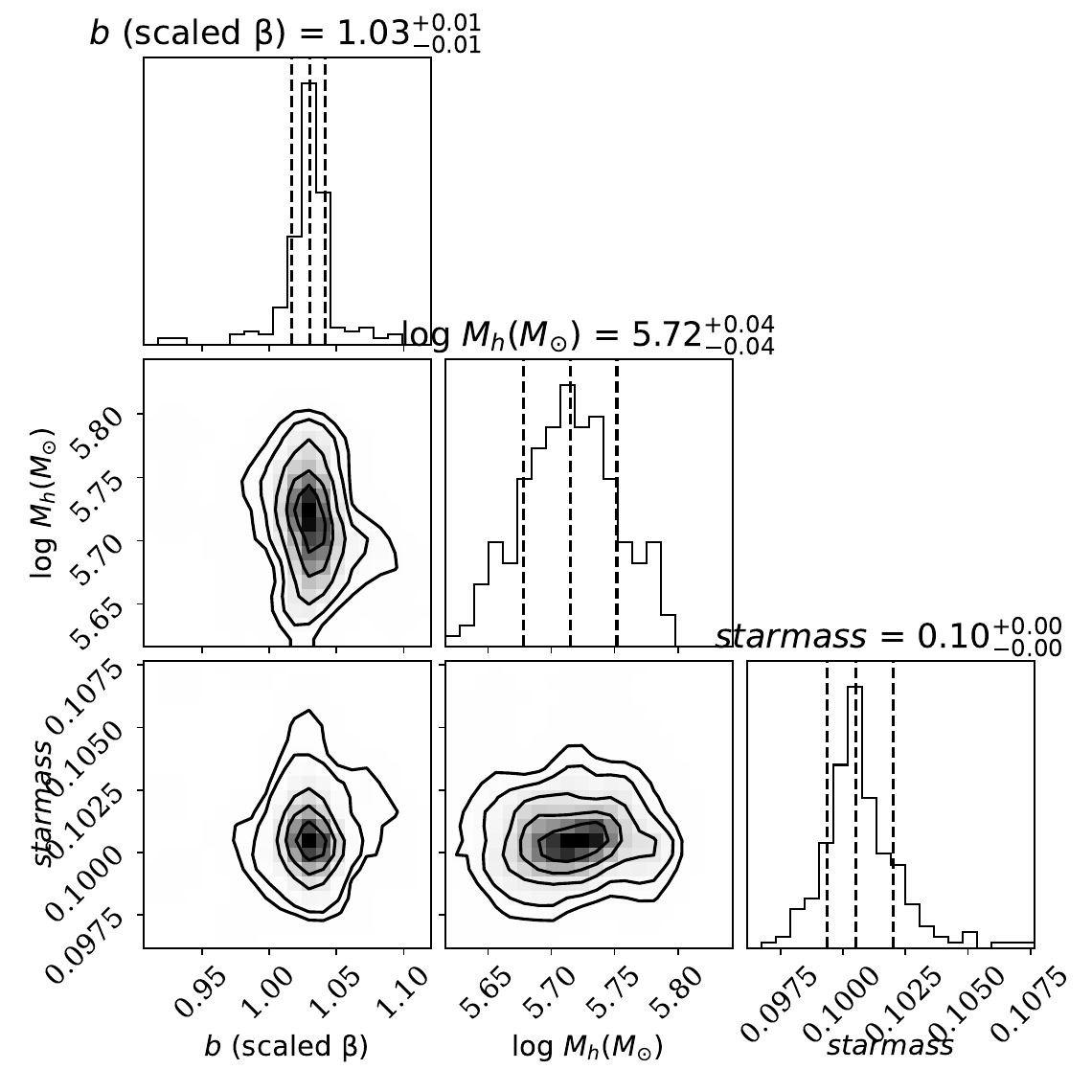}
    \caption{Posterior distribution of the fitting parameter: $M_{\rm BH}$, $m_{*}$ and scaled penetration factor $b$. The four panels correspond to light curves with pre-peak data points from ASASS-SN $g$ band only (AS, top left), ATLAS $c$ band only (ATc, top right), ASAS-SN $g$ band $+$ ATLAS $c$ band (AS+ATc, bottom left), ASAS-SN $g$ band $+$ ATLAS $c$, $o$ band (AS+ATco, bottom right). The veritcal dashed lines indicate the 14th, 50th and 86th percentiles of the distribution. We take the 50th percentile as the best-fit results, the 14th and 86th percentiles indicate the $1\sigma$ uncertainty.}
    \label{fig:corner}
\end{figure}

% Table 1
\begin{table}[htbp]
\caption{The fitting parameters used in the light curve modeling.
\label{table-paras}}
\begin{center}
\begin{tabular}{cccc}
  \hline
  Parameter & Prior Distribution Type & Min  &  Max   \\
  \hline
  $M_{\rm BH}/M_{\odot}$  & Log    & $10^5$  & $10^7$ \\
  $m_{*}/M_{\odot}$       & Kroupa & $0.08$ & $3$ \\
  $b$ (scaled penetration factor)  & Flat & $0$ & $2$ \\
  $\eta$                  & Flat   & $10^{-4}$ & $0.4$ \\
  $R_{\rm ph0}$           & Log    & $10^{-4}$ & $10^4$ \\
  $l$                     & Flat   & $0$ & $4$ \\
  $t_{\nu}$/days          & Log    & $10^{-3}$ & $10^5$ \\
  \hline
\end{tabular}
\end{center}
\textbf{NOTES:}~The first column gives the name of the parameters. The second column indicate the type of prior distribution for each parameters: ``Flat" means the prior is uniformly sampled from the value range; ``Log" means the prior is logrithmically uniformly sampled in the value range; ``Kroupa" means the stellar mass are sampled from the Kroupa initial mass function. The third and fourth columns give the allowed range for each parameter.
\end{table}

% Table 2
\begin{table}[htbp]
\caption{The fitting results and derived quantities for different composite light curves.}
\label{table-results}
\begin{center}
\begin{tabular}{c|ccc|ccc|c}
  \hline
  Pre-peak data source & $\log(M_{\rm BH}/M_{\odot})$ & $m_{*}/M_{\odot}$  &  $b$ & $\log(R_I/\rm{cm})$ & $\eta_{\rm sh}$  & $\eta_{\rm fit}$ & WAIC \\
  \hline
  AS      & $5.67^{+0.05}_{-0.06}$ & $0.10^{+0.01}_{-0.00}$  & $1.16^{+0.04}_{-0.04}$ & $13.78$ & $1.70\times 10^{-3}$ & $2.63\times 10^{-4}$ &$559.1$\\
  ATc     & $5.82^{+0.05}_{-0.04}$ & $0.99^{+0.02}_{-0.03}$  & $0.93^{+0.03}_{-0.04}$ & $14.42$ & $3.13\times 10^{-4}$ & $1.05\times 10^{-4}$ &$581.4$\\
  AS+ATc  & $5.81^{+0.04}_{-0.05}$ & $0.98^{+0.02}_{-0.04}$  & $0.93^{+0.02}_{-0.03}$ & $14.42$ & $2.99\times 10^{-4}$ & $1.05\times 10^{-4}$ &$575.0$\\
  AS+ATco & $5.72^{+0.04}_{-0.04}$ & $0.10^{+0.00}_{-0.00}$  & $1.03^{+0.01}_{-0.01}$ & $13.94$ & $1.18\times 10^{-3}$ & $2.75\times 10^{-4}$ &$530.6$\\
  \hline
\end{tabular}
\end{center}
\textbf{NOTES:}~First column indicate the data source of the pre-peak light curves, see definition in the text (Section~\ref{SUBSECT:LC_OTHERS}). From second to fourth columns are the black hole mass, stellar mass and the scaled penetration factor, respectively. The fifth and sixth columns are the radial distance of debris self-intersection point and the shock dissipation efficiency (equation~\ref{Eq:eta_sh}), respectively, calculated from the median values of the relevant fitting parameters. The seventh column is the median radiation efficiency from the fitting. The eighth column is the WAIC score, introduced in Section~\ref{SUBSEC:parameter}.
\end{table}

% Table 3
\begin{table}[htbp]
\caption{The timescales for different composite light curves.}
\label{table-timescales}
\begin{center}
\begin{tabular}{c|ccc}
  \hline
  Pre-peak data source & $t_{\rm 1/2}$~/~day & $t_{\rm mb}$~/~day & $\log(t_{\nu}$~/~day) \\
  \hline
  AS      & $10.76^{+1.42}_{-1.38}$ & $12.83^{+0.97}_{-0.88}$ & $0.21^{+0.16}_{-0.54}$\\
  ATc     & $ 7.09^{+1.12}_{-0.90}$ & $28.00^{+1.57}_{-1.43}$ & $-0.63^{+0.73}_{-1.33}$\\
  AS+ATc  & $ 6.96^{+1.26}_{-0.93}$ & $27.47^{+1.61}_{-1.49}$ & $-0.46^{+0.53}_{-1.34}$\\
  AS+ATco & $10.17^{+0.87}_{-0.86}$ & $13.43^{+0.72}_{-0.66}$ & $0.16^{+0.10}_{-0.17}$\\
  \hline
\end{tabular}
\end{center}
\textbf{NOTES:} First column indicate the data source of the pre-peak light curves, see definition in the text (Section~\ref{SUBSECT:LC_OTHERS}). Second column is the time from half peak luminosity to the peak luminosity measured from the mock light curves generated by \texttt{MOSFiT}, which is used as an estimation of the rising timescale of the light curves. Third column is the orbital period for the most tightly bound debris, derived from $a_{\rm mb}$ (equation 4 of \cite{Dai+2015ApJ}) assuming the debris moves on Keplerian orbit. $a_{\rm mb}$ depends on $M_{\rm BH}$, $m_*$ and $r_*$. The former two quantities are fitting parameters and their values are drawn from their posterior distributions, then $r_*$ is calculated from the \citet{Tout+1996MNRAS} stellar mass-radius relation implemented in \texttt{MOSFiT}. Fourth column is the viscous timescales. As in Figure~\ref{fig:corner}, we use the 14th and 86th percentiles to indicate the $1\sigma$ uncertainty in all three timescales.
\end{table}

%%%%%%%%%%%%%%%%%%%%%%%%%%%%%%%%%%%%%%%%%%%%%%%%%%%%%%%%%%%%%%%%
%%%%%%%%%%%%%%%%%%%%%%%%%%%%%%%%%%%%%%%%%%%%%%%%%%%%%%%%%%%%%%%%
\section{Results and Discussion}
\label{SECT:Result_Discussion}

\subsection{Fitting Results}
\label{SUBSEC:results}
The mock light curves generated by the fitting results and the observed data points are plotted in Figure~\ref{fig:lightcurve}. The black hole mass, stellar mass and scaled penetration factor $b$ are listed in Table~\ref{table-results}.

The fitting results reported from the four composite light curves falls into two cases: ($a$) full disruption ($b>1$) of a low mass star ($\sim 0.1 M_{\odot}$); ($b$) partial disruption ($b<1$) of a near solar mass star. Besides, the rising time of case ($a$) is longer than that of case ($b$) (see column 2 of Table~\ref{table-timescales}), and the black hole mass in case ($a$) is slightly lower than in case ($b$). Nevertheless, all the black hole masses are lower than that reported by \cite{ZJW2023}. Besides, we note there is some inconsistency between ASAS-SN and ATLAS data. In the cases where the rising part only contains one data source (e.g. AS and ATc), the rising part is well fitted (the observed data point falls on the mock light curves). While in the cases where the rising part adopt data from both ASAS-SN and ATLAS (e.g. AS+ATc and AS+ATco), the rising part is not well fitted.

Recently some interesting TDEs have been observed: their light curves show re-brightening features in the declining phase after the first peak. The separation between the first and second peaks spans a range from a few hundred days (in the sample of \cite{Yao+2023ApJ}) to a few years \citep{Somalwar2023arXiv}. The nature of this phenomenon is still an open question, there are some possible models: repeated partial TDEs produced by the same star \citep{Somalwar2023arXiv}; two subsequent TDEs produced by two different stars following the tidal breakup of a binary \citep{Mandel_Levin2015ApJ}.

In the case of AT 2023clx, the scaled penetration factor $b$ obtained from the four composite light curves are slightly different: while the composite light curve (AS, AS+ATco) reports a complete disruption of the intruding star ($b>1$), the other two light curves (ATc, AS+ATc) report partial disruptions ($b<1$). However, in the latter cases, $b$ is very close to 1, thus the survived remnant core is unlikely to produce a second TDE, given its extremely low mass (hence the tidal radius is very small) and the possibility of ejection from the SMBH (see for example, \cite{Zhong+2022ApJ}). In either case, we speculate that there would be no further TDE flare from this galaxy, until another star falls into the tidal radius.

Although \texttt{MOSFiT} has been successful in fitting the UV/optical light curve of many TDEs, there are still some limitations for this code and may affect our measurements.

The radiation efficiency $\eta$ is fixed throughout the lifetime of a TDE. However, in the long term evolution, as the new dissipation processes add in, $\eta$ shall vary with time (see Section~\ref{SUBSEC:energy_source}). Therefore, \texttt{MOSFiT} should be applied to a limited segment of the light curve in which the $\eta$ value can be taken as a constant, but it is hard to find out this time segment in advance. The mock light curves in the decline phase are generally well matched with the observation data, hence we think a fixed $\eta$ value is appropriate in the fitting procedure, although some small deviations are noticeable.

\cite{MR2021ApJ} found that there is a degeneracy between $\eta$ and $m_*$ in the fitting results of \texttt{MOSFiT}. They reduced the lower limit of the $\eta$ prior distribution and re-fitted the TDEs presented in \cite{MGR2019}. They found that for the same TDE light curve, lowering the $\eta$ value by a factor of 10 will result in stellar mass that is roughly 10 times heavier. Our fitting results also show that lower $\eta_{\rm fit}$ corresponds to higher $m_*$, however, we argue that this behavior is not due to the parameter degeneracy. The difference among the four fitted $m_*$ is $\sim 10$, then the difference of $\eta_{\rm fit}$ caused by the parameter degeneracy should also be $\sim 10$, but the results only show a factor of $\sim 2.7$ difference.

The viscous timescale (Column 4 in Table~\ref{table-timescales}) is much shorter than the fallback time (Column 3 in Table~\ref{table-timescales}), indicating that the bolometric light curve well represents the mass fallback rate. Therefore, the accuracy of estimations on the $M_{\rm BH}$, $m_*$ and $\beta$ depends on the mass fallback rate model. \texttt{MOSFiT} adopts the mass fallback rate obtained from the disruption of stars modeled by polytropes \citep{GRR2013}. The $\gamma=5/3$ polytrope is a good approximation for a $m_*\simeq 0.1 M_{\odot}$ star (obtained from the AS, AS+ATco light curves), and the stellar structure hardly changes within the age of the Universe. But for the $m_*\simeq 1 M_{\odot}$ star (obtained from the ATc, AS+ATc light curves), the stellar structure becomes more centrally concentrated than the $\gamma=4/3$ polytrope as it ages, and the fallback rate shall be modified~\citep{Law-Smith+2020}. The age of the disrupted star could not be obtained from the light curve alone, since there is a degeneracy with $\beta$~\citep{Law-Smith+2020}. The chemical abundance of the tidal debris is promising in determining the age of the disrupted star~\citep{Law-Smith+2019}. Unfortunately, we do not have spectrum data to study the age and/or type of the disrupted star. The parameters fitted from the ATc, AS+ATc light curves should be treated with caution, although they have higher WAIC scores.

%%%%%%%%%%%%%%%%%%%%%%%%%%%%%%%%%%%%%%%%%%%%%%
\subsection{What Process Powers the Optical Radiation of AT 2023clx}
\label{SUBSEC:energy_source}

After a star is disrupted, the debris bound to the SMBH is not immediately accreted. Instead, the debris shall spend some time to settle into an accretion disk. The specific orbital energy of the bound debris is spread between $-GM_{\rm BH}r_{*}/r_{\rm t}^2$ and $0$ \citep{EK1989}. Therefore, different portion of the debris returns to the pericenter at different times. Due to the general relativistic apsidal precession of the orbit, the outgoing stream is deflected toward the fallback stream and shall collide with it at some place. Such collision could dissipate the orbital energy of the debris, and promote the formation of accretion disk \citep{Kim_Park_Lee1999ApJ,Jiang_2016ApJ}. In the classic picture of TDE, it is assumed that the debris can quickly form a compact circular accretion disk (radial size $\sim 2r_{\rm p}$), and the spectrum energy distribution (SED) of this disk peaks at soft X-ray or EUV band \citep{Rees1988}. The accretion rate at the early stage shall exceed the Eddington accretion rate for $M_{\rm BH} < 10^7 M_{\odot}$, hence the disk could launch radiation driven outflows, which obscures the central accretion disk and forms a reprocessing layer (RL) that convert the X-ray/EUV photons into UV/optical photons \citep{SQ2009,MS2016MNRAS}. However, the assumption that an accretion disk is quickly formed has been put into question by the later numerical and theoretical works \citep{Shiokawa+2015ApJ,Piran+2015ApJ}, because the strength of energy dissipation during the stream-stream collision is too mild to circularize the debris in a short period of time.

\texttt{MOSFiT} does not assume any specific optical radiation mechanism. Instead, it adopts a dynamical photosphere whose size evolves with the bolometric luminosity. In addition, as mentioned by ~\cite{MGR2019}, \texttt{MOSFiT} allows a wide range of radiation efficiency, hence it could cover various of radiation models, including the accretion disk, and the stream-stream collision. Note, in the case of orbital energy being dissipated in the stream-stream collision, the $\dot{M}_{\rm acc}$ shall be substituted with $\dot{M}_{\rm inflow}$ and interpreted as the mass inflow rate to the collision region (see for example, \cite{Jiang_2016ApJ}). Based on the fitting results, we discuss the possible energy source of AT 2023clx.

The $b$ values indicate that the penetration factor $\beta$ is very close to $\beta_{\rm d}$. From these parameters, we find the pericenter distance $r_{\rm p}$ is much larger than  the Schwarzschild radius of the central SMBH. Therefore, the general relativistic apsidal precession effect on the trajectories of the tidal debris is mild. According the self-intersection model of \cite{Dai+2015ApJ}, we find in AT 2023clx the stream-stream collision happens at a place far from the SMBH (order of $10^{13}$--$10^{14}$ cm, see the fifth column of Table~\ref{table-results}, for comparison, the pericenter distance is order of $10^{12}$ cm). The shocks in the collision region dissipate the orbital energy into thermal energy and part of it may eventually radiate away. The amount of dissipated energy can be estimated with the semimajor axis of the debris before and after the collision,

\begin{equation}
\Delta E = r_{\rm g} \left ( \frac{1}{a_{\rm pc}} - \frac{1}{a_{\rm mb}}\right )mc^2,
\label{Eq:delta_E}
\end{equation}
\noindent
where $a_{\rm mb}$ is the semimajor axis of the most bound debris (equation 4 of \cite{Dai+2015ApJ}) and $a_{\rm pc}$ is the post-collision semimajor axis of that debris (equation 11 of \cite{Dai+2015ApJ}). This dissipated energy set the upper limit for the radiation luminosity powered by stream-stream collision, and we define the corresponding efficiency as

\begin{equation}
\eta_{\rm sh} \equiv \frac{\Delta E}{mc^2} = \frac{r_{\rm g}(a_{\rm mb} - a_{\rm pc})}{a_{\rm mb} a_{\rm pc}},
\label{Eq:eta_sh}
\end{equation}
\noindent
In the case of AT 2023clx, the maximum $\eta_{\rm sh}$ value is a few times $10^{-3}$ (see the sixth column of Table~\ref{table-results}). The radiation efficiency reported from all the fittings (the seventh column of Table~\ref{table-results}) are smaller than the $\eta_{\rm sh}$ values derived above, and also much lower than the typical radiation efficiency of an standard accretion disk ($\sim 0.1$). The rest of the dissipated energy may turn into the kinetic energy of a mild collision-induced outflow (CIO) \citep{Lu_Bonnerot2020MNRAS,Ryu+2023ApJ.957.12R}.

Note, in the stream-stream collision scenario, the optical photons are not directly coming from the collision region, since the temperature of the shock region and the post-shock gas could reach $10^6$ K \citep{Ryu+2023ApJ.957.12R}. Instead, they are generated in the photosphere embedded in the CIO~\citep{Jiang_2016ApJ}.

So far, we only considered the energy dissipation in the first collision. The post-collision debris still moves on highly eccentric orbit, and might experience more collisions. \cite{Bonnerot+2017MNRAS} and \cite{Chen_Dou_Shen2022ApJ} have analytically studied the long term evolution of the debris stream. They find that the successive collisions should dissipate more orbital energy, hence the $\eta_{\rm sh}$ should evolve with time, a feature not captured by \texttt{MOSFiT}. The calculations of \cite{Bonnerot+2017MNRAS} and \cite{Chen_Dou_Shen2022ApJ} assume the debris strictly follows the trajectory modeled by succession of elliptical orbits with decreasing orbital energy (and angular momentum if magnetic stress is applied) and relativistic precessed pericenters. However, numerical simulations show that after the first stream-stream collision, the post-collision streams may not follow the theoretical elliptical trajectory, due to the redistribution of angular momentum by the shocks~\citep{Shiokawa+2015ApJ}. Instead, the post-collision stream quickly form an eccentric disk, whose inner edge is still far from the accretion radius of the black hole. The shocks formed inside the eccentric disk could dissipate more orbital energy into heat, although the efficiency is low ~\citep{Shiokawa+2015ApJ,Piran+2015ApJ}.

On the other hand, \citet{LCA2021} proposed that the optical emission from an eccentric disk is powered by accretion onto the black hole, instead of the self-crossing shocks. In this model, the viscous friction at the disk pericenter region works as a new heating source. The disk is geometric thin and optically thick and the emission region for optical photons lies in the disk surface around apocenter region. \cite{ZLK2021} has estimated the total radiation efficiency of this viscous dissipation process and the stream-stream collision process. According to our fitted black hole mass, stellar mass and penetration factor, we read from the Fig. 1 of \cite{ZLK2021} that the total radiation efficiency is around $\log(\eta)\approx -2.5$, which brings the luminosity to sub-Eddington, hence no strong outflow is expected. This $\eta$ is higher than our fitting results, sufficient to power the optical luminosity. But there is no outflow or disk expansion to absorb the rest of the dissipated energy, this model may overestimate the dissipated energy and is disfavored.

The low efficiency could also due to the low black hole mass. \citet{Davis_Laor2011} (hereafter DL11) reported a relation between accretion efficiency and black hole mass, $\eta_{\rm DL11} = 0.089 (M_{\rm BH}/10^8 M_{\odot})^{0.52}$, from 80 Palomar-Green quasars. They defined the accretion efficiency as $\eta_{\rm DL11}=L_{\rm bol}/(\dot{M}_{\rm BH}c^2)$, where $L_{\rm bol}$ is the bolometric luminosity integrated from infrared to X-ray band, and $\dot{M}_{\rm BH}$ is derived from the UV/optical SED based on fully relativistic thin accretion disk model \citep{Dexter_Agol2009}. When disk wind is taken into account, the normalization of the above relation shall be reduced by roughly 0.5 dex, according to Figure 10 of \cite{Laor_Davis2014}. In \texttt{MOSFiT}, $\eta$ solely accounts for the bolometric luminosity of optical bands. Hence the $\eta_{\rm DL11}$ value derived from the DL11 relation shall be further reduced by roughly 1 dex (see Figure 12 of DL11), in order to compare with our results. Inserting our fitted black hole mass into the DL11 relation, and applying the corrections caused by disk wind and the different definitions of bolometric luminosity, we find $\eta_{\rm DL11}$ is lower than $\eta_{\rm fit}$. In addition, the aforementioned works use thin accretion disk model in their calculation, which is generally not realized in the TDE accretion flow. Therefore, we think the DL11 relation is not likely responsible for the low $\eta_{\rm fit}$ value. We also notice that there are suspicions on the DL11 relation in the literature: it may be an artifact of selection effect \citep{Raimundo+2012} and the uncertainty in the black hole mass estimation \citep{Wu+2013}.

The fallback rate could exceed Eddington accretion rate for smaller black holes. In the context of standard accretion disk with $\eta\sim 0.1$, the radiation driven outflow could reduce the actual accretion rate to $f_{\rm acc}\dot{M}_{\rm fb}$, with $f_{\rm acc}\ll 1$ \citep{SQ2009,MS2016MNRAS}. In this case, the $\eta_{\rm fit}$ reported by \texttt{MOSFiT} is actually $\eta f_{\rm acc}$ ($\ll 0.1$). We think this situation is not likely occurred in AT 2023clx, because (1) the time needed for forming a standard accretion disk is longer than our observation time (see Section~\ref{SUBSEC:X_ray}); (2) the peak bolometric luminosity is only about $0.1 L_{\rm Edd}$~\citep{ZJW2023}.

In summary, the low $\eta_{\rm fit}$ values support the conjecture that the optical luminosity is powered by the energy released in the stream-stream collision. Another clue comes from the viscous timescale $t_{\nu}$: in all the fittings $t_{\nu}$ is much smaller than the fallback timescale, so that the light curve closely follow the mass fallback rate, which is inline with the explanation that the luminosity (around the peak) is powered by stream-stream collision.

%%%%%%%%%%%%%%%%%%%%%%%%%%%%%%%%%%%%%%%%%%%%%%
\subsection{Reason for the Non-detection in Soft X-ray Band}
\label{SUBSEC:X_ray}

\cite{ZJW2023} reports non-detection of soft X-ray photons with \textit{Swift}/XRT observation on AT 2023clx within 90 days after the optical peak. According to the discussion in the previous paragraphs, and assume the soft X-ray photons are solely emitted from the compact accretion disk (because the soft X-ray SEDs from other non-jetted TDEs indicate a thermal-origin, \cite{Saxton+2020SSRv}), we conjecture that the non-detection is simply because the compact accretion disk is yet to form. In fact, many optical TDEs exhibit later brightening of soft X-ray emissions long after the optical peak, e.g., ASASSN-15oi \citep{Gezari+2017ApJL,Holoien+2018MNRAS}, AT2019azh \citep{vanVelzen2021ApJ,LDC+2022ApJ}, OGLE16aaa \citep{Kajava+2020AA}. In long term evolution, the accretion flow eventually settles down into a compact standard circular accretion disk.

Numerical simulations suggest that the circularization happens roughly $\sim 10 t_{\rm mb}$ after the disruption, where $t_{\rm mb}$ is the orbital period of the most bound debris \citep{Shiokawa+2015ApJ}. More specifically, we estimate the time to circularization ($t_{\rm circ}$) using the dimensionless disk formation efficiency $\mathcal{C}(M_{\rm BH},m_*,\beta)$ defined by \cite{Wong+2022ApJL}, which is the ratio of the energy loss during the first stream-stream collision (Equation~\ref{Eq:delta_E}) to the total energy that need to be dissipated before the circularization finishes. Then the quantity $1/\mathcal{C}$ gives a rough estimate on the number of collisions needed to circularize the debris, hence $t_{\rm circ} \sim t_{\rm mb}/\mathcal{C}$. The values of $t_{\rm mb}$ derived from our fitting results are presented in Table~\ref{table-timescales}. For the parameters obtained from the AS and AS+ATco light curves, $t_{\rm mb} \simeq 13$ days and $\mathcal{C}\simeq0.129$, thus it takes $\sim 100$ days to form the compact accretion disk. While for the parameters obtained from the ATc and AS+ATc light curves, $t_{\rm mb} \simeq 28$ days and $\mathcal{C}\simeq0.042$, the timescale extends to $\sim 600$ days. Until then may the X-ray photons emitted from the inner disk region (near the innermost stable circular orbit, ISCO) been observed. Our estimated timescales are roughly consistent with the duration of non-detection (within 90 days after optical peak).

The non-detection of X-ray photons could also be understood in the framework of the TDE unified model~\citep{Dai+2018ApJ}. This model employs an outflow-launching super-Eddington accretion disk, and the outflow plays the role of reprocessing layer that absorbs the soft X-ray photons emitted from the accretion disk and re-emits in the UV/optical band. The geometric shape of the reprocessing layer is non-spherical ($R_{\rm ph}$ varies with inclination angle), hence the soft X-ray to optical flux ratio varies with the inclination angle $i$ between the line-of-sight and the rotation axis of the accretion disk. \cite{Thomsen+2022ApJ} studied the SEDs of this model with different accretion rates, and investigated the temporal evolution of post-peak black body temperature ($T_{\rm O,BB}$) inferred from optical emission, as well as the optical to soft X-ray flux ratio viewed from different inclination angle at different times. In the case of AT 2023clx, the post-peak $T_{\rm O,BB}$ increases slowly with time (as discussed in section~\ref{SUBSEC:LightCurveModel}, this can be inferred from the value of $l$, all four composite light curves report $l$ values between $0.7$ and $1.1$). According to \cite{Thomsen+2022ApJ}, such temporal evolution of $T_{\rm O,BB}$ can be reproduced in the moderate inclination angle cases (e.g. $i=30^{\circ}$ and $i=50^{\circ}$). For these inclination angles, their calculations show that as the accretion rate declines, the soft X-ray flux that can be observed increases (Fig. 4 of~\cite{Thomsen+2022ApJ}), because the density of the outflow falls down.

The TDE unified model assumes a compact circular accretion disk as the source of the outflow and the soft X-ray photons. A swift formation of such disk is feasible only if $\beta>3$ when $M_{\rm BH} < 5\times 10^6 M_{\odot}$~\citep{Dai+2015ApJ}. However, as we have shown, it takes $100$ to $600$ days to form such a compact circular accretion disk. In addition, the mass fallback rate might falls below the Eddington accretion rate at that time. Hence, it is unlikely that the outflow launched from the super-Eddington accretion disk is responsible for the optical emission and the non-detection of X-ray photons in the early stage.

Another problem for the formation of a compact circular disk comes from the magnetic stress operating in the stream-stream collision and subsequent accretion process, since every disrupted star should possesses magnetic field. Magnetic stress can transport the angular momentum of the stream outward \citep{Svirski+2017MNRAS}. \cite{Bonnerot+2017MNRAS} found that in the case of strong magnetic stress, the post-collision stream may lose angular momentum significantly, and result in ballistic accretion in a short period of time, when pericenter of the stream falls below the radius of unstable circular orbit. As a result, a circular accretion disk will never form. Unfortunately, our observation data can not constrain the role of magnetic stress in AT 2023clx.

We encourage continuously monitoring the nucleus of NGC 3799 in the soft X-ray band to further test our conjecture.

%%%%%%%%%%%%%%%%%%%%%%%%%%%%%%%%%%%%%%%%%%%%%%%%%%%%%%%%%%%%%%%%
%%%%%%%%%%%%%%%%%%%%%%%%%%%%%%%%%%%%%%%%%%%%%%%%%%%%%%%%%%%%%%%%
\section{Summary}
\label{SECT:Summary}
In this paper, we present the optical light curves of AT 2023clx in the declining phase, observed with Mephisto in the $uvgr$ bands. Combining our light curves with the ASAS-SN and ATLAS data in the rising phase, we obtained the full composite light curves. Then we use the light curve fitting software \texttt{MOSFiT} to extract the physical paramters for this particular TDE, especially the black hole mass, the stellar mass and the penetration factor. We constructed four groups of composite light curves, each uses different data in the rising phase. The four fitted black hole masses are close to each other ($10^{5.67}$--$10^{5.82}~M_{\odot}$), but they are all lower than the estimation made from the SMBH-galaxy mass relation ($10^{6.26 \pm 0.28}~M_{\odot}$, \cite{ZJW2023}).

The other parameters are clearly divided into two categories: either a full disruption of a low mass star, or a partial disruption of a near solar mass star (Section~\ref{SUBSEC:results}). This discrepancy is mainly caused by the inconsistency between the rising phase ASAS-SN and ATLAS data, which shows the importance of obtaining good measurements in the rising phase. If the rising phase is well sampled and measured (for example the full light curve is obtained with Mephisto), the fitting result would be improved.

The origin of the optical emissions in TDEs is still an open question \citep{Gezari2021ARAA}. The major origins of the optical photons include: (1) reprocessing layer embedded in the outflow launched from the accretion disk during super-Eddington accretion phase \citep{SQ2009,MS2016MNRAS}; (2) photosphere embedded in the collision-induced outflow launched from the location of stream-stream collision \citep{Jiang_2016ApJ,Lu_Bonnerot2020MNRAS,Ryu+2023ApJ.957.12R}; (3) the surface layer in the apocenter portion of an eccentric accretion disk \citep{LCA2021}.
Using the physical parameters obtained from the light curve fitting, we explain that the observed optical emission (i.e. before MJD 60100) should being powered by the stream-stream collision process and emanated from the photosphere embedded in the CIO, due to the low radiation efficiency and also the long timescale needed to circularize the debris, hence unlikely to launch disk outflow that made the reprocessing layer. And because during the observation campaign, a compact accretion disk is not formed, hence the emission in the soft X-ray should be weak or even not produced. The structure of the debris flow shall evolve with time, and the other two mechanisms may take over the stream-stream collision and become the main engine of the optical luminosity in the future. But we do not have the chance to test them for AT 2023clx in long term evolution, because it is the faintest TDE observed to date \citep{ZJW2023}, and the optical luminosity already falls below the host galaxy background. However, we still have a chance to catch the soft X-ray photons when the debris circularized into an compact accretion disk.

The stream-stream collision and the CIO could explain the optical emissions of AT 2023clx, we note that this model still has caveat. The high temperature collision region is enshrouded by the CIO, causing difficulty in producing thermal soft X-ray flux~\citep{Jiang_2016ApJ}. Therefore, it can not be applied to the X-ray-selected TDEs, and the TDEs that have comparable X-ray and optical fluxes. For such TDEs, the TDE unified model \citep{Dai+2018ApJ,Thomsen+2022ApJ} is more feasible.

\normalem
\begin{acknowledgements}
Mephisto is developed at and operated by the South-Western Institute for Astronomy Research of Yunnan University (SWIFAR-YNU), funded by the “Yunnan University Development Plan for WorldClass University” and “Yunnan University Development Plan for World-Class Astronomy Discipline”. The authors acknowledge supports from the “Science \& Technology Champion Project” (202005AB160002) and from two “Team Projects” – the “Innovation Team” (202105AE160021) and the “Top Team” (202305AT350002), all funded by the “Yunnan Revitalization Talent Support Program”. 

\end{acknowledgements}
  
\bibliographystyle{raa}
\bibliography{ms2024-0095}

\begin{thebibliography}{62}
\providecommand\natexlab[1]{#1}
\providecommand\JournalTitle[1]{#1}

\bibitem[{Bellm} {et~al.}(2019)]{Bellm2019PASP..131a8002B}
{Bellm}, E.~C., {Kulkarni}, S.~R., {Graham}, M.~J., {et~al.} 2019, \pasp, 131, 018002

\bibitem[{Bonnerot} {et~al.}(2017)]{Bonnerot+2017MNRAS}
{Bonnerot}, C., {Rossi}, E.~M., \& {Lodato}, G. 2017, \mnras, 464, 2816

\bibitem[{Chen} {et~al.}(2022)]{Chen_Dou_Shen2022ApJ}
{Chen}, J.-H., {Dou}, L.-M., \& {Shen}, R.-F. 2022, \apj, 928, 63

\bibitem[{Dai} {et~al.}(2015)]{Dai+2015ApJ}
{Dai}, L., {McKinney}, J.~C., \& {Miller}, M.~C. 2015, \apjl, 812, L39

\bibitem[{Dai} {et~al.}(2018)]{Dai+2018ApJ}
{Dai}, L., {McKinney}, J.~C., {Roth}, N., {Ramirez-Ruiz}, E., \& {Miller}, M.~C. 2018, \apjl, 859, L20

\bibitem[{Davis} \& {Laor}(2011)]{Davis_Laor2011}
{Davis}, S.~W., \& {Laor}, A. 2011, \apj, 728, 98

\bibitem[{Dexter} \& {Agol}(2009)]{Dexter_Agol2009}
{Dexter}, J., \& {Agol}, E. 2009, \apj, 696, 1616

\bibitem[{Evans} \& {Kochanek}(1989)]{EK1989}
{Evans}, C.~R., \& {Kochanek}, C.~S. 1989, \apjl, 346, L13

\bibitem[{Gezari}(2021)]{Gezari2021ARAA}
{Gezari}, S. 2021, \araa, 59, 21

\bibitem[{Gezari} {et~al.}(2017)]{Gezari+2017ApJL}
{Gezari}, S., {Cenko}, S.~B., \& {Arcavi}, I. 2017, \apjl, 851, L47

\bibitem[{Gomez} {et~al.}(2020)]{Gomez+2020}
{Gomez}, S., {Nicholl}, M., {Short}, P., {et~al.} 2020, \mnras, 497, 1925

\bibitem[{Graham} {et~al.}(2019)]{Graham2019PASP}
{Graham}, M.~J., {Kulkarni}, S.~R., {Bellm}, E.~C., {et~al.} 2019, \pasp, 131, 078001

\bibitem[{Guillochon} {et~al.}(2018)]{MOSFiT}
{Guillochon}, J., {Nicholl}, M., {Villar}, V.~A., {et~al.} 2018, \apjs, 236, 6

\bibitem[{Guillochon} \& {Ramirez-Ruiz}(2013)]{GRR2013}
{Guillochon}, J., \& {Ramirez-Ruiz}, E. 2013, \apj, 767, 25

\bibitem[{Hammerstein} {et~al.}(2023)]{HvVC2023}
{Hammerstein}, E., {van Velzen}, S., {Gezari}, S., {et~al.} 2023, \apj, 942, 9

\bibitem[{Holoien} {et~al.}(2018)]{Holoien+2018MNRAS}
{Holoien}, T.~W.~S., {Brown}, J.~S., {Auchettl}, K., {et~al.} 2018, \mnras, 480, 5689

\bibitem[{Ivezi{\'c}} {et~al.}(2019)]{Ivezic2019ApJ...873..111I}
{Ivezi{\'c}}, {\v{Z}}., {Kahn}, S.~M., {Tyson}, J.~A., {et~al.} 2019, \apj, 873, 111

\bibitem[{Jiang} {et~al.}(2016)]{Jiang_2016ApJ}
{Jiang}, Y.-F., {Guillochon}, J., \& {Loeb}, A. 2016, \apj, 830, 125

\bibitem[{Kajava} {et~al.}(2020)]{Kajava+2020AA}
{Kajava}, J. J.~E., {Giustini}, M., {Saxton}, R.~D., \& {Miniutti}, G. 2020, \aap, 639, A100

\bibitem[{Kim} {et~al.}(1999)]{Kim_Park_Lee1999ApJ}
{Kim}, S.~S., {Park}, M.-G., \& {Lee}, H.~M. 1999, \apj, 519, 647

\bibitem[{Kochanek} {et~al.}(2017)]{KSS2017}
{Kochanek}, C.~S., {Shappee}, B.~J., {Stanek}, K.~Z., {et~al.} 2017, \pasp, 129, 104502

\bibitem[{Kov{\'a}cs-Stermeczky} \& {Vink{\'o}}(2023{\natexlab{a}})]{TiDE}
{Kov{\'a}cs-Stermeczky}, Z.~V., \& {Vink{\'o}}, J. 2023{\natexlab{a}}, \pasp, 135, 034102

\bibitem[{Kov{\'a}cs-Stermeczky} \& {Vink{\'o}}(2023{\natexlab{b}})]{KV2023}
{Kov{\'a}cs-Stermeczky}, Z.~V., \& {Vink{\'o}}, J. 2023{\natexlab{b}}, \pasp, 135, 104102

\bibitem[{Kroupa} {et~al.}(1993)]{Kroupa+1993MNRAS}
{Kroupa}, P., {Tout}, C.~A., \& {Gilmore}, G. 1993, \mnras, 262, 545

\bibitem[{Laor} \& {Davis}(2014)]{Laor_Davis2014}
{Laor}, A., \& {Davis}, S.~W. 2014, \mnras, 438, 3024

\bibitem[{Law-Smith} {et~al.}(2020)]{Law-Smith+2020}
{Law-Smith}, J. A.~P., {Coulter}, D.~A., {Guillochon}, J., {Mockler}, B., \& {Ramirez-Ruiz}, E. 2020, \apj, 905, 141

\bibitem[{Law-Smith} {et~al.}(2019)]{Law-Smith+2019}
{Law-Smith}, J., {Guillochon}, J., \& {Ramirez-Ruiz}, E. 2019, \apjl, 882, L25

\bibitem[{Lin} {et~al.}(2018)]{Lin+2018NatAs}
{Lin}, D., {Strader}, J., {Carrasco}, E.~R., {et~al.} 2018, Nature Astronomy, 2, 656

\bibitem[{Liu} {et~al.}(2021)]{LCA2021}
{Liu}, F.~K., {Cao}, C.~Y., {Abramowicz}, M.~A., {et~al.} 2021, \apj, 908, 179

\bibitem[{Liu} {et~al.}(2022)]{LDC+2022ApJ}
{Liu}, X.-L., {Dou}, L.-M., {Chen}, J.-H., \& {Shen}, R.-F. 2022, \apj, 925, 67

\bibitem[{Lu} \& {Bonnerot}(2020)]{Lu_Bonnerot2020MNRAS}
{Lu}, W., \& {Bonnerot}, C. 2020, \mnras, 492, 686

\bibitem[{Mandel} \& {Levin}(2015)]{Mandel_Levin2015ApJ}
{Mandel}, I., \& {Levin}, Y. 2015, \apjl, 805, L4

\bibitem[{Metzger} \& {Stone}(2016)]{MS2016MNRAS}
{Metzger}, B.~D., \& {Stone}, N.~C. 2016, \mnras, 461, 948

\bibitem[{Mockler} {et~al.}(2019)]{MGR2019}
{Mockler}, B., {Guillochon}, J., \& {Ramirez-Ruiz}, E. 2019, \apj, 872, 151

\bibitem[{Mockler} \& {Ramirez-Ruiz}(2021)]{MR2021ApJ}
{Mockler}, B., \& {Ramirez-Ruiz}, E. 2021, \apj, 906, 101

\bibitem[{Nicholl} {et~al.}(2020)]{Nicholl+2020MNRAS}
{Nicholl}, M., {Wevers}, T., {Oates}, S.~R., {et~al.} 2020, \mnras, 499, 482

\bibitem[{Piran} {et~al.}(2015)]{Piran+2015ApJ}
{Piran}, T., {Svirski}, G., {Krolik}, J., {Cheng}, R.~M., \& {Shiokawa}, H. 2015, \apj, 806, 164

\bibitem[{Raimundo} {et~al.}(2012)]{Raimundo+2012}
{Raimundo}, S.~I., {Fabian}, A.~C., {Vasudevan}, R.~V., {Gandhi}, P., \& {Wu}, J. 2012, \mnras, 419, 2529

\bibitem[{Rees}(1988)]{Rees1988}
{Rees}, M.~J. 1988, \nat, 333, 523

\bibitem[{Reines} \& {Volonteri}(2015)]{Reines_Volonteri_2015ApJ}
{Reines}, A.~E., \& {Volonteri}, M. 2015, \apj, 813, 82

\bibitem[{Ryu} {et~al.}(2020)]{TDEmass}
{Ryu}, T., {Krolik}, J., \& {Piran}, T. 2020, \apj, 904, 73

\bibitem[{Ryu} {et~al.}(2023)]{Ryu+2023ApJ.957.12R}
{Ryu}, T., {Krolik}, J., {Piran}, T., {Noble}, S.~C., \& {Avara}, M. 2023, \apj, 957, 12

\bibitem[{Saxton} {et~al.}(2020)]{Saxton+2020SSRv}
{Saxton}, R., {Komossa}, S., {Auchettl}, K., \& {Jonker}, P.~G. 2020, \ssr, 216, 85

\bibitem[{Schlafly} \& {Finkbeiner}(2011)]{SF2011ApJ}
{Schlafly}, E.~F., \& {Finkbeiner}, D.~P. 2011, \apj, 737, 103

\bibitem[{Schlegel} {et~al.}(1998)]{SFD1998ApJ}
{Schlegel}, D.~J., {Finkbeiner}, D.~P., \& {Davis}, M. 1998, \apj, 500, 525

\bibitem[{Shappee} {et~al.}(2014)]{SPG2014}
{Shappee}, B.~J., {Prieto}, J.~L., {Grupe}, D., {et~al.} 2014, \apj, 788, 48

\bibitem[{Shiokawa} {et~al.}(2015)]{Shiokawa+2015ApJ}
{Shiokawa}, H., {Krolik}, J.~H., {Cheng}, R.~M., {Piran}, T., \& {Noble}, S.~C. 2015, \apj, 804, 85

\bibitem[{Smith} {et~al.}(2020)]{SSY2020PASP}
{Smith}, K.~W., {Smartt}, S.~J., {Young}, D.~R., {et~al.} 2020, \pasp, 132, 085002

\bibitem[{Somalwar} {et~al.}(2023)]{Somalwar2023arXiv}
{Somalwar}, J.~J., {Ravi}, V., {Yao}, Y., {et~al.} 2023, arXiv e-prints, arXiv:2310.03782

\bibitem[{Strubbe} \& {Quataert}(2009)]{SQ2009}
{Strubbe}, L.~E., \& {Quataert}, E. 2009, \mnras, 400, 2070

\bibitem[{Svirski} {et~al.}(2017)]{Svirski+2017MNRAS}
{Svirski}, G., {Piran}, T., \& {Krolik}, J. 2017, \mnras, 467, 1426

\bibitem[{Taguchi} {et~al.}(2023)]{Taguchi2023}
{Taguchi}, K., {Uno}, K., {Nagao}, T., \& {Maeda}, K. 2023, Transient Name Server Classification Report, 2023-438, 1

\bibitem[{Thomsen} {et~al.}(2022)]{Thomsen+2022ApJ}
{Thomsen}, L.~L., {Kwan}, T.~M., {Dai}, L., {et~al.} 2022, \apjl, 937, L28

\bibitem[{Tonry} {et~al.}(2018)]{TDH2018PASP}
{Tonry}, J.~L., {Denneau}, L., {Heinze}, A.~N., {et~al.} 2018, \pasp, 130, 064505

\bibitem[{Tout} {et~al.}(1996)]{Tout+1996MNRAS}
{Tout}, C.~A., {Pols}, O.~R., {Eggleton}, P.~P., \& {Han}, Z. 1996, \mnras, 281, 257

\bibitem[{van Velzen} {et~al.}(2021)]{vanVelzen2021ApJ}
{van Velzen}, S., {Gezari}, S., {Hammerstein}, E., {et~al.} 2021, \apj, 908, 4

\bibitem[{Wong} {et~al.}(2022)]{Wong+2022ApJL}
{Wong}, T. H.~T., {Pfister}, H., \& {Dai}, L. 2022, \apjl, 927, L19

\bibitem[{Wu} {et~al.}(2013)]{Wu+2013}
{Wu}, S., {Lu}, Y., {Zhang}, F., \& {Lu}, Y. 2013, \mnras, 436, 3271

\bibitem[{Yao} {et~al.}(2023)]{Yao+2023ApJ}
{Yao}, Y., {Ravi}, V., {Gezari}, S., {et~al.} 2023, \apjl, 955, L6

\bibitem[{Zhong} {et~al.}(2022)]{Zhong+2022ApJ}
{Zhong}, S., {Li}, S., {Berczik}, P., \& {Spurzem}, R. 2022, \apj, 933, 96

\bibitem[{Zhou} {et~al.}(2021)]{ZLK2021}
{Zhou}, Z.~Q., {Liu}, F.~K., {Komossa}, S., {et~al.} 2021, \apj, 907, 77

\bibitem[{Zhu} {et~al.}(2023)]{ZJW2023}
{Zhu}, J., {Jiang}, N., {Wang}, T., {et~al.} 2023, \apjl, 952, L35

\end{thebibliography}

\end{document}